\documentclass[sigconf]{acmart}
\setcopyright{none}

\copyrightyear{2025}
\acmYear{2025}
\acmConference[SIGIR '25]{Proceedings of the 48th International ACM SIGIR Conference on Research and Development in Information Retrieval}{July 13--18, 2025}{Padova, Italy}
\acmBooktitle{Proceedings of the 48th International ACM SIGIR Conference on Research and Development in Information Retrieval (SIGIR '25), July 13--18, 2025, Padova, Italy}


\usepackage{multirow}
\usepackage{enumitem}
\usepackage{balance}
\usepackage{hhline}
\usepackage{makecell}

\usepackage[capitalize]{cleveref}
\crefname{section}{Sec.}{Secs.}
\Crefname{section}{Section}{Sections}
\Crefname{table}{Table}{Tables}
\crefname{table}{Tab.}{Tabs.}
\crefname{definition}{Def.}{Defs.}
\Crefname{definition}{Definition}{Definitions}

\newcommand\best[1]{\textbf{#1}}    
\newcommand\scnd[1]{\underline{#1}}  
\author{Chanwoo Kim}
\email{chanwoo.kim@snu.ac.kr}
\affiliation{
  \institution{Seoul National Univ.}
  \city{Seoul}
  \country{Korea}
}

\author{Jinkyu Sung}
\email{jinkyusung@snu.ac.kr}
\affiliation{
  \institution{Seoul National Univ.}
  \city{Seoul}
  \country{Korea}
}

\author{Yebonn Han}
\email{bonnida@snu.ac.kr}
\affiliation{
  \institution{Seoul National Univ.}
  \city{Seoul}
  \country{Korea}
}

\author{Joonseok Lee}
\email{joonseok@snu.ac.kr}
\authornote{Corresponding author}
\affiliation{
  \institution{Seoul National Univ.}
  \city{Seoul}
  \country{Korea}
}
\title{Graph Spectral Filtering with Chebyshev Interpolation for Recommendation}

\begin{abstract}
Graph convolutional networks have recently gained prominence in collaborative filtering (CF) for recommendations. However, we identify potential bottlenecks in two foundational components. First, the embedding layer leads to a latent space with limited capacity, overlooking locally observed but potentially valuable preference patterns. Also, the widely-used neighborhood aggregation is limited in its ability to leverage diverse preference patterns in a fine-grained manner. Building on spectral graph theory, we reveal that these limitations stem from graph filtering with a cut-off in the frequency spectrum and a restricted linear form. To address these issues, we introduce ChebyCF, a CF framework based on graph spectral filtering. Instead of a learned embedding, it takes a user's raw interaction history to utilize the full spectrum of signals contained in it. Also, it adopts Chebyshev interpolation to effectively approximate a flexible non-linear graph filter, and further enhances it by using an additional ideal pass filter and degree-based normalization. Through extensive experiments, we verify that ChebyCF overcomes the aforementioned bottlenecks and achieves state-of-the-art performance across multiple benchmarks and reasonably fast inference. Our code is available at \url{https://github.com/chanwoo0806/ChebyCF}.
\end{abstract}

\begin{document}
\maketitle

\section{Introduction}

Recommender systems play a pivotal role in modern commercial services with the increased number and diversity of options.
Among the various recommendation methodologies, collaborative filtering

\newpage

\noindent
(CF) \cite{mnihProbabilisticMatrixFactorization2007, rendleBPRBayesianPersonalized2009, korenMatrixFactorizationTechniques2009, rendleFactorizationMachines2010} leverages preference patterns within user-item interaction data to recommend similar items to users who share similar tastes.
Thanks to its major advantage of enabling effective personalization without the help of additional information about users or items, CF remains as a fundamental approach in recommendations \cite{sedhainAutoRecAutoencodersMeet2015, covingtonDeepNeuralNetworks2016, heNeuralCollaborativeFiltering2017, liangVariationalAutoencodersCollaborative2018, steckEmbarrassinglyShallowAutoencoders2019, maoSimpleXSimpleStrong2021, wangRepresentationAlignmentUniformity2022}.

Recently, CF approaches treating the interaction data as a graph demonstrate state-of-the-art performance.
Equipped with the capability to identify preference patterns that can only be captured through high-order connectivity within the graph, graph-based CF models effectively mitigate the limitations of traditional CF models, which rely only on sparse signals from direct interactions between a user and an item.
In particular, with the advances of graph convolutional networks (GCNs) \cite{kipfSemiSupervisedClassificationGraph2017, velickovicGraphAttentionNetworks2017}, GCN-based CF has emerged as a dominant paradigm.

While the initial GCN-based approaches \cite{bergGraphConvolutionalMatrix2018, yingGraphConvolutionalNeural2018, wangNeuralGraphCollaborative2019} have simply applied the vanilla GCNs to the recommendation, numerous subsequent research \cite{sunNeighborInteractionAware2020, wangDisentangledGraphCollaborative2020, liuInterestawareMessagePassingGCN2021, fanGraphTrendFiltering2022, kongLinearNonLinearThat2022, wangCollaborationAwareGraphConvolutional2023} adapt them to the CF-based recommendation, considering that the original GCNs have been developed for different graph problems, \emph{e.g.}, node classification.
For example, it has been proposed to remove complex node-wise feature transformations, as user and item features are absent in CF scenarios \cite{chenRevisitingGraphBased2020, heLightGCNSimplifyingPowering2020, maoUltraGCNUltraSimplification2021}.
Self-supervised graph learning also has been incorporated to mitigate the sparsity of interaction data \cite{wuSelfsupervisedGraphLearning2021, linImprovingGraphCollaborative2022, xiaHypergraphContrastiveCollaborative2022, jiangAdaptiveGraphContrastive2023, renSSLRecSelfSupervisedLearning2024}.

However, we question two fundamental elements that were established during the early stages of GCN-based CF research and have been used as a default without reexamination.
First, inherited from the traditional CF models, the embedding layer confines the preference signals into a \textit{latent space with limited capacity}.
Optimized to learn the most globally prominent signals from the graph, the model loses the opportunity to take advantage of locally observed patterns, carrying over the same limitation of the earlier CF models \cite{leeLocalLowRankMatrix2013}.
Second, in most cases, the graph structure is utilized only by the \textit{simple neighborhood aggregation}.
Unlike the traditional GCN tasks \cite{kipfSemiSupervisedClassificationGraph2017} where abundant node features are available, the graph (interaction data) serves as the sole information in CF tasks, making it essential to make the most of the graph information.
For this reason, we hypothesize that a more sophisticated method of leveraging diverse patterns in the graph would be necessary.
We then formalize these two limitations of GCN-based CF through the lens of spectral graph theory \cite{shumanEmergingFieldSignal2013, ortegaGraphSignalProcessing2018}, demonstrating that they arise from two key constraints in the graph filtering process: the inability to utilize the full frequency spectrum and the restriction to a linear form.

In this paper, we propose a solution named ChebyCF, based on graph spectral filtering (GSF).
Built on the item-item graph, ChebyCF directly takes the users' whole interactions as input graph signal, eliminating the need for embedding users and items into a latent space, which is limited in its ability to accommodate entire frequency spectrum.
Also, our ChebyCF employs Chebyshev interpolation \cite{defferrardConvolutionalNeuralNetworks2016, chenRevisitingGraphBased2020} to effectively approximate a flexible non-linear graph filter without requiring the graph Laplacian matrix decomposition, and achieves nuanced filtering by adopting the plateau function as the approximation target.
Additionally, we address the limitations of smooth graph filters and popularity bias by incorporating an ideal pass filter \cite{shenHowPowerfulGraph2021} and degree-based normalization \cite{liuPersonalizedGraphSignal2023}, respectively.
Through extensive experiments, we demonstrate that ChebyCF effectively overcomes the aforementioned bottlenecks and achieves the state-of-the-art performance.
Notably, ChebyCF's lightweight yet precise filtering enables it to surpass the performance of the existing top-performing CF models, while achieving up to 8.1x faster inference than the previous best model.

Our contributions can be summarized as follows:  
\begin{itemize}[leftmargin=5mm]
    \item We identify two key limitations of GCN-based CF from the perspective of spectral graph theory: 1) challenges in utilizing the full frequency spectrum and 2) overly inflexible filtering.
    \item We propose ChebyCF, a novel approach that adopts graph spectral filtering, effectively approximating a flexible non-linear graph filter leveraging Chebyshev interpolation.
    \item We experimentally verify that ChebyCF achieves the state-of-the-art in recommendation, along with fast inference time.
\end{itemize}
\section{Preliminaries}
\label{sec:prelim}

We formulate the problem and introduce spectral graph theory \cite{shumanEmergingFieldSignal2013, ortegaGraphSignalProcessing2018} to form the foundation for the discussions in this paper.
In addition, we revisit how GCNs have been applied to CF, following representative research \cite{kipfSemiSupervisedClassificationGraph2017, wuSimplifyingGraphConvolutional2019, chenRevisitingGraphBased2020, heLightGCNSimplifyingPowering2020}.

\subsection{Problem Formulation}
\label{sec:problem}

We aim to solve the recommendation problem with implicit feedback.
Given the sets of users and items, denoted by $\mathcal{U}$ and $\mathcal{I}$ respectively, and the interaction matrix $\mathbf{R} \in \{0,1\}^{|\mathcal{U}| \times |\mathcal{I}|}$ with $\mathbf{R}_{ui} = 1$ if user $u$ has interacted with item $i$ and $\mathbf{R}_{ui} = 0$ otherwise, the goal is to estimate the preference matrix $\mathbf{\hat{R}} \in \mathbb{R}^{|\mathcal{U}| \times |\mathcal{I}|}$, where $\mathbf{\hat{R}}_{ui}$ represents the predicted score of the user $u$'s preference on item $i$.

\subsection{Spectral Graph Theory}
\label{sec:prelim:graph_theory}

\textbf{Graph Laplacian.}
Let $\mathcal{G} = (\mathcal{V}, \mathcal{E})$ be a weighted undirected graph with $n$ vertices, where $\mathcal{V}$ and $\mathcal{E}$ denote the vertex set and edge set, respectively.
The graph can be represented either by an adjacency matrix $\mathbf{A} \in \mathbb{R}^{n \times n}$ or by a normalized graph Laplacian matrix $\mathbf{L} = \mathbf{I} - \mathbf{D}^{-1/2} \mathbf{A} \mathbf{D}^{-1/2}$, where $\mathbf{I}$ is the identity matrix, $\mathbf{D} = \text{diag}(\mathbf{A}\mathbf{1})$ is the degree matrix, and $\mathbf{1}$ is the all-ones vector.
Since $\mathbf{L}$ is a real symmetric matrix, it can be eigendecomposed to $\mathbf{L} = \mathbf{Q} \boldsymbol{\Lambda} \mathbf{Q}^\top$, where $\boldsymbol{\Lambda} = \text{diag}(\lambda_1, \dots, \lambda_n)$ is the diagonal matrix of eigenvalues and $\mathbf{Q}$ is the matrix of orthonormal eigenvectors.

\vspace{0.1cm} \noindent
\textbf{Graph Fourier Transform.}
A graph signal is denoted as $\mathbf{x} \in \mathbb{R}^n$, where the $i$-th element $\mathbf{x}_i$ corresponds to the scalar value at $i$-th vertex.
The graph Fourier transform (GFT) maps a graph signal from the spatial (vertex) domain to the spectral (frequency) domain by interpreting the eigenvectors of the graph Laplacian as basis signals and their corresponding eigenvalues as frequencies.
Low-frequency components reflect smooth variations across the graph, whereas high-frequency ones capture rapid and localized oscillations.
Specifically, the smoothness of a graph signal is defined by the similarity of values across adjacent nodes; formally, $S(\mathbf{x}) = \sum_{(i,j) \in \mathcal{E}} \mathbf{A}_{ij} (\mathbf{x}_i - \mathbf{x}_j)^2 = \mathbf{x}^\top\mathbf{L}\mathbf{x}$.
The smoothness $S(\mathbf{q})$ of an eigenvector $\mathbf{q}$ is equal to its associated eigenvalue $\lambda$.
GFT and its inverse GFT are computed as $\hat{\mathbf{x}} = \mathbf{Q}^\top \mathbf{x}$ and $\mathbf{x} = \mathbf{Q} \hat{\mathbf{x}}$, respectively.

\vspace{0.1cm} \noindent
\textbf{Graph Spectral Filtering.}
Graph spectral filtering involves transforming a graph signal into the spectral domain with GFT, adjusting the signals based on a transfer function, and returning back to the spatial domain by inverse GFT.
Specifically, the transfer function $h(\lambda)$ reduces or amplifies the contribution of the basis signal corresponding to frequency $\lambda$.
Formally, a graph filter $H$ is defined as follows.
\begin{definition}[Graph Filter]
  Given a transfer function $h(\lambda)$, a graph Laplacian matrix $\mathbf{L}$ and its eigenvectors $\mathbf{Q}$, a graph filter $H(\mathbf{L}) \in \mathbb{R}^{n\times n}$ is defined as
  \begin{equation}
    H(\mathbf{L}) = \mathbf{Q} \text{diag}(h(\lambda_1), \dots, h(\lambda_n)) \mathbf{Q}^\top.
  \end{equation}
\end{definition}
\noindent Given an input graph signal $\mathbf{x} \in \mathbb{R}^n$, the filtered signal is expressed as $\mathbf{y} = H(\mathbf{L})\mathbf{x}$.

\begin{figure}
    \centering
    \includegraphics[width=\linewidth]{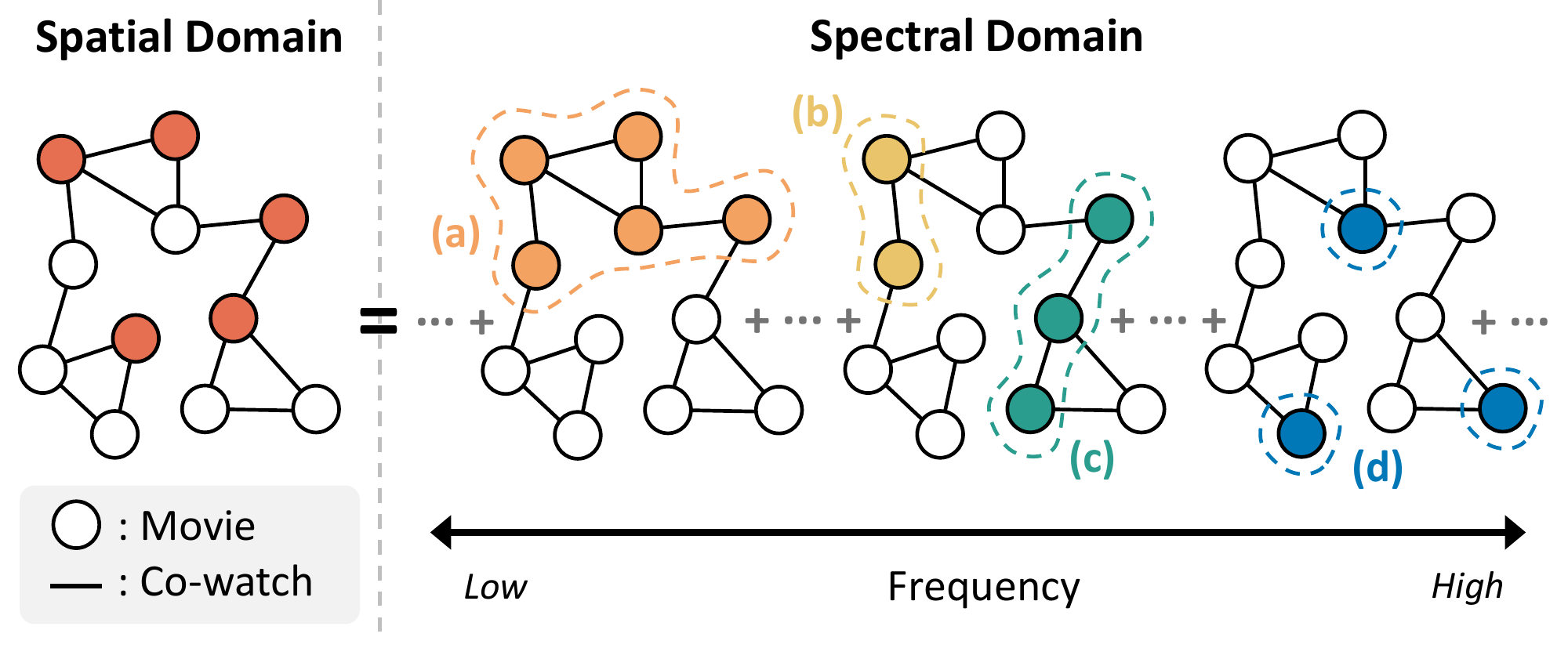}
    \vspace{-20pt}
    \caption{Graph spectral view in movie recommendation. Red denotes movies watched by the user.}
    \label{fig:preliminary}
    \vspace{-0.2cm}
\end{figure}

\vspace{0.1cm} \noindent
\textbf{Application.}
We illustrate the application using movie recommendation as an example.
Consider an item-item graph, where each node represents a movie and the edge weights indicate the extent to which two movies are frequently co-watched.
A user's watch history corresponds to a graph signal, and can be decomposed into signals of various frequencies, as illustrated in \cref{fig:preliminary}.
Low-frequency signals are composed of movies commonly watched together by many users (\cref{fig:preliminary} (a)), whereas high-frequency signals contain movies that are typically not co-watched (\cref{fig:preliminary} (d)).

The philosophy of CF is to leverage the watch histories of many users collaboratively to extract shared patterns as indicators of preference, while tends to disregard unshared patterns, which are frequently attributed to noises (\textit{e.g.}, misclicks).
Hence, low-pass graph filtering, which emphasizes low-frequency signals over high-frequency signals, is well-suited for CF-based recommendation.
However, designing a filter that assigns an optimal weight to each frequency component is a nontrivial task.
For instance, mid-frequency signals, consisting of locally co-watched movies (\cref{fig:preliminary} (b,c)), may not capture globally dominant patterns but still hold potential for fine-grained personalization.
Thus, careful weighting would be crucial to fully exploit them.

\subsection{GCN-based CF}
\label{sec:prelim:gcn_for_recs}

In GCNs, a standard graph convolutional layer performs two primary operations: 1) neighborhood aggregation, which updates node features by leveraging the graph structure, and 2) node-wise feature transformation via a multi-layer perceptron (MLP).
Specifically, at the $l$-th layer, it performs
\begin{equation}
    \label{eq:gcn}
    \mathbf{X}^{(l)} = \text{MLP}_l ( \tilde{\mathbf{A}}\mathbf{X}^{(l-1)} ),
\end{equation}
where $\tilde{\mathbf{A}}$ is the normalized adjacency matrix with self-loops, $\mathbf{X}^{(0)}$ is input node features, and $\mathbf{X}^{(l)}$ is the node feature matrix after the $l$-th layer for $l = 1, ..., L$.
Neighborhood aggregation indicates multiplication of $\tilde{\mathbf{A}}$, as it averages the feature information of neighboring nodes.

In most GCN-based CF settings, the graph is defined with users and items as nodes, and interactions as edges. The normalized interaction matrix is defined as $\tilde{\mathbf{R}} = \mathbf{D}_\mathcal{U}^{-1/2} \mathbf{R} \mathbf{D}_\mathcal{I}^{-1/2}$, where $\mathbf{D}_\mathcal{U} = \text{diag}(\mathbf{R}\mathbf{1})$ and $\mathbf{D}_\mathcal{I} = \text{diag}(\mathbf{1}^T \mathbf{R})$ are the degree matrices of users and items, respectively.
Based on this, the adjacency matrix is defined as
\begin{equation}
    \label{eq:adj}
    \mathbf{A}^{\ast} = \begin{bmatrix} 0 & \tilde{\mathbf{R}} \\ \tilde{\mathbf{R}}^\top & 0 \end{bmatrix}.
\end{equation}
In the CF setting where no side information about the users and items is available, the input node features are replaced with  a learnable embedding matrix $\mathbf{E} \in \mathbb{R}^{(|\mathcal{U}| + |\mathcal{I}|) \times d}$.
In short, the elements in Eq.~\eqref{eq:gcn} are adapted as follows:
\begin{equation}
\label{eq:gcn-cf}
    \mathbf{X}^{(0)} \overset{(i)}{=} \mathbf{E}
    \quad \text{and} \quad
    \mathbf{X}^{(l)} \overset{(ii)}{=}
    \mathbf{A}^{\ast}\mathbf{X}^{(l-1)}.
\end{equation}
\section{Motivation}
\label{sec:motiv}

In spite of the success of GCN-based CF models in recommendations, however, we claim that they are still not optimally adapted to the recommendation task.
Particularly, the following two components remain as bottlenecks: 1) the user and item \textit{embeddings with limited capacity}, and 2) overly simple form of \textit{neighborhood aggregation}.

\vspace{0.2cm} \noindent
\textbf{Embedding Space with Limited Capacity.}
An embedding layer is introduced to map each user and item to a $d$-dimensional latent space (Eq.~\eqref{eq:gcn-cf}-(i)).
This is based on the long-lasting low-rank assumption of the interaction matrix in recommendation literature, where we believe there are up to $d$ important factors ruling the preference patterns.
This restricted space may significantly prevent the model from finding all useful patterns in the interaction matrix, when it is far beyond the assumed low rank $d$.

A series of seminal papers \cite{leeLocalLowRankMatrix2013,leeLocalCollaborativeRanking2014,lee2016llorma,choiLocalCollaborativeAutoencoders2021}, in fact, have demonstrated that the ground-truth preference matrix is only locally low-rank within a coherent set of users and items, rather than globally low-rank.
These studies have achieved superior recommendation performance under this local low-rank assumption, leveraging locally discovered patterns.
In that sense, we similarly hypothesize that the embedding layer in GCN-based may overlook local preference patterns in the graph, which are potentially important.

\begin{figure}
    \centering
    \includegraphics[height=4.2cm]{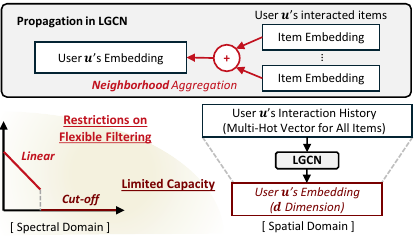}
    \vspace{-5pt}
    \caption{
    In the spectral view, LGCN's neighborhood aggregation with an embedding layer corresponds to linear low-pass filtering with a cutoff.}
    \label{fig:motiv}
    \vspace{-0.3cm}
\end{figure}

\vspace{0.2cm} \noindent
\textbf{Simplistic Neighborhood Aggregation.}
In the graph convolutional layer, the graph structure is mainly leveraged by the neighborhood aggregation, which is often fixed as a simple weighted averaging of the neighboring node features.
Complex operations are concentrated in the subsequent MLP, where features are transformed independently at each node.
In a subsequent study \cite{wuSimplifyingGraphConvolutional2019}, GCNs have been even re-interpreted as MLPs accompanied by graph-based preprocessing.

In the CF setting, however, input node features are absent, and complex feature transformations are also being omitted \cite{chenRevisitingGraphBased2020, heLightGCNSimplifyingPowering2020, maoUltraGCNUltraSimplification2021, fanGraphTrendFiltering2022}, while the neighborhood aggregation remains intact (Eq.~\eqref{eq:gcn-cf}-(ii)).
Since the graph (interaction data) is the sole input, we hypothesize that relying on the simple neighborhood aggregation would be insufficient and a more sophisticated operation should be designed to fully leverage diverse preference patterns within the interactions.

\vspace{0.2cm} \noindent
\textbf{Analysis from Graph Spectral Perspective.}
We further analyze GCN-based CF from the perspective of spectral graph theory in \cref{sec:prelim:graph_theory}. 
Specifically, under the assumption that the embeddings maximally preserve the interaction data, we claim that neighborhood aggregation with an embedding layer is equivalent to graph spectral filtering based on a \textit{linear low-pass filter} with a \textit{cutoff}.
The cut-off arises from the $d$-dimensional constraint of the embeddings, where the information beyond the $d$ lowest frequencies is discarded, making higher-frequency information inaccessible, and the linearity arises from the neighborhood aggregation, restricting the adjustment of each frequency component to a simple linear way.
This is illustrated in \cref{fig:motiv} and formally described in the following theorem.

\noindent
\begin{theorem}
\label{theorem:bottleneck}

Consider the \textit{Linear Graph Convolutional Network} (LGCN) \citep{heLightGCNSimplifyingPowering2020}, where propagation is performed via the neighborhood aggregation as in \cref{eq:gcn-cf}-(ii), and
the embedding matrices $\hat{\mathbf{E}}_\mathcal{U} \in \mathbb{R}^{\mathcal{|U|}  \times d }$ and $\hat{\mathbf{E}}_\mathcal{I} \in \mathbb{R}^{\mathcal{|I|}  \times d }$ are set by solving
\begin{align*}
  \left( \hat{\mathbf{E}}_\mathcal{U}, \hat{\mathbf{E}}_\mathcal{I} \right) &= \underset{(\mathbf{E}_\mathcal{U}, \mathbf{E}_\mathcal{I})}{\mathrm{argmin}}\ \| \Tilde{\mathbf{R}} - \mathbf{E}_\mathcal{U} \mathbf{E}_\mathcal{I}^\top \|_F 
  \ \  \text{s.t.}  \ \ 
  \mathrm{rank}(\mathbf{E}_\mathcal{U}),\mathrm{rank}(\mathbf{E}_\mathcal{I}) \le d,
\end{align*}
with $d \le \min(|\mathcal{U}|, |\mathcal{I}|)$ as the embedding dimension and $\Tilde{\mathbf{R}}$ as the normalized interaction matrix.
Then, $\hat{\mathbf{r}}_u \in \mathbb{R}^{|\mathcal{I}|}$, the predicted preference of user $u$, is equivalent to the spectrally filtered output of the input signal $\Tilde{\mathbf{r}}_u$ (\textit{i.e.}, the $u$-th row of $\Tilde{\mathbf{R}}$) over the normalized item-item graph with Laplacian $\mathbf{L}^\ast = \mathbf{I} - \Tilde{\mathbf{R}}^{\top}\Tilde{\mathbf{R}}$, using the transfer function $h:[0,1] \to [0,1]$ defined by
\[
h(\lambda) = \underbrace{(1-\lambda)}_{\text{Linear Low-pass}} \cdot \quad \underbrace{\mathbb{I}[\lambda \le \lambda_d]}_{\text{Cut-off}},
\]
where $\mathbb{I}$ is the indicator function and $\lambda_d$ is the $d$-th smallest frequency.
\end{theorem}

Removing frequencies higher than the cut-off would not pose any issue if they solely contain unnecessary noise (analogous to the low-rank assumption).
However, as discussed earlier, this assumption does not hold well.
Furthermore, since the only available data in CF is the graph, it is crucial to optimally weight graph signals across diverse frequencies to fully exploit the underlying preference patterns.
In this context, the constraint of linearity can act as a significant impediment.
\section{Methodology}
\label{sec:method}

To address the posed bottlenecks of GCN-based CF identified in \cref{sec:motiv}, we propose a novel graph filtering called ChebyCF, which leverages signals across the entire frequency spectrum and filters them in a refined manner that goes beyond linear.

\subsection{Graph and Signal}
\label{sec:method:graph}

We adopt the item-item graph structure, where nodes represent items and edges correspond to normalized co-occurrence, measured by the number of users who have interacted with both items.
Formally, the graph Laplacian is given by $\mathbf{L^\ast} = \mathbf{I} - \Tilde{\mathbf{R}}^{\top}\Tilde{\mathbf{R}}$, the same as in \cref{theorem:bottleneck}.
For the graph signal of a user $u$, we take the $u$-th row of $\mathbf{R}$, denoted by $\mathbf{r}_u \in \mathbb{R}^{|\mathcal{I}|}$. 
One notable aspect that distinguishes our method from previous works is that the users are directly represented with their interactions with items on demand, instead of mapping them to a $d$-dimensional parametric embedding, which is limited in its ability to capture full frequency spectrum.

\subsection{Chebyshev Filter}
\label{sec:method:cheby_filter}

To enable more flexible filtering beyond linear low-pass filters, we extend the form of the transfer function for graph filters.
While the most flexible form involves analytic functions, it requires the full eigendecomposition of the graph Laplacian, which entails a computational complexity of $O(n^3)$ with $n$ denoting the number of the nodes.
This is impractical for real-world recommender systems with a large number of items.
However, the eigendecomposition can be avoided by constraining the transfer function to a polynomial form, allowing filtering to be performed using only matrix polynomials of the graph Laplacian.

\begin{figure}
    \centering
    \includegraphics[width=0.9\linewidth]{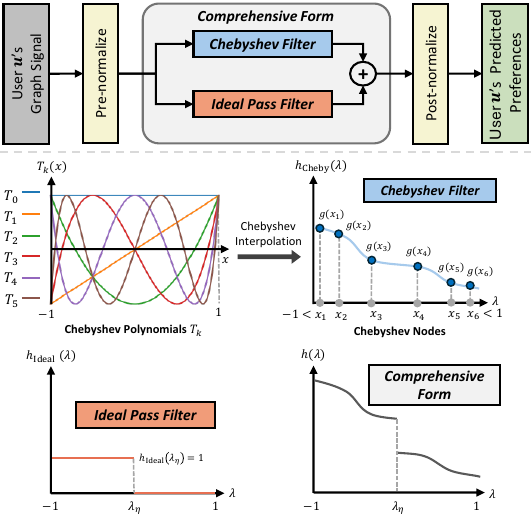}
    \vspace{-0.4cm}
    \caption{An illustration of the proposed ChebyCF.}
    \label{fig:method}
    \vspace{-0.4cm}
\end{figure}

To achieve both flexibility and computational efficiency, we employ the Chebyshev polynomials (\cref{fig:method} top-left), whose truncated expansions are known to provide a minimax polynomial approximation for analytic functions \cite{geddes1978near}.
Specifically, we adopt Chebyshev interpolation, which uses Chebyshev nodes as interpolation points. Since the Chebyshev nodes are clustered more densely near the boundaries (\cref{fig:method} top-right), they mitigate the Runge phenomenon, which is a common issue of severe oscillations near the boundaries when using equispaced points \cite{heConvolutionalNeuralNetworks2022}.
In this way, we achieve a graph filter that offers a superior approximation of analytic functions while remaining matrix decomposition-free.

\begin{definition}[Chebyshev Polynomials]
    \label{def:chebypoly}
    The Chebyshev polynomials of the first kind are defined through the recurrence relation:
        \begin{equation}
            T_k(x) = 2xT_{k-1}(x) - T_{k-2}(x), \quad \forall k \geq 2 
        \end{equation}
    with initial conditions $T_0(x) = 1$ and $T_1(x) = x$.
\end{definition}

\begin{definition}[Chebyshev Nodes]
    The $i$-th Chebyshev nodes for $T_k$ are given by
    \begin{equation}
        x_i = \cos\left(\frac{2k+1-2i}{2k} \pi\right), \quad \forall i \in \{1, \dots, k\}.
    \end{equation}
    The nodes, which are the $k$ roots of $T_k$, lie in the interval $(-1, 1)$.
\end{definition}

\begin{definition}[Chebyshev Interpolation]
    \label{def:chebyinter}
    Let $\{x_i\}_{i=1}^{K+1}$ be the Chebyshev nodes of $T_{K+1}$. When a function $g$ is approximated with a $K$-th order Chebyshev expansion, the coefficient $c_k$ of $T_k$ is given by 
    \begin{equation}
        c_k = \frac{a_k}{K + 1} \sum_{i=1}^{K+1} g(x_i) T_k(x_i), \quad \forall k \in \{0, \dots, K\},
    \end{equation}
    where $a_k = 1$ for $k = 0$, and $a_k = 2$ otherwise.
\end{definition}

Accordingly, we formally define our $K$-th order transfer function as $h_{\text{Cheby}}(\lambda)=\sum_{k=0}^{K}c_kT_k(\lambda)$ and corresponding Chebyshev filter
\begin{equation}
  \label{eq:cheby_filter}
  H_{\text{Cheby}}(\mathbf{L^\ast}) = \sum_{k=0}^{K}  c_k T_k(\Tilde{\mathbf{L}}),
\end{equation}
where $c_k$ is the coefficient in \cref{def:chebyinter}, $g$ in $c_k$ denotes the target transfer function to be approximated, and ${\Tilde{\mathbf{L}}} = 2\mathbf{L^\ast} - \mathbf{I}$ is the rescaled Laplacian matrix.
Since the largest eigenvalue of $\mathbf{L^\ast}$ is equal to $1$ as proven in Theorem 4.1 in \cite{shenHowPowerfulGraph2021}, the rescaling maps the eigenvalues to the interval $[-1, 1]$, the domain where Chebyshev interpolation is ensured to make a stable approximation, while keeping the eigenvectors intact.

As the target transfer function $g$ can be freely designed to perform the desirable recommendation, we suggest one modest option motivated by prior observations.
We adopt the plateau function, which not only exhibits a low-pass filtering as discussed in \cref{sec:prelim:graph_theory}, but also offers additional flexibility by enabling control over width of the spectral extremes that should be treated as essential low or noisy high frequencies, through the degree of flatness ($\phi$):
\begin{equation}
    \label{eq:plateau}
    g_{\text{plateau}}(\lambda) =
    \begin{cases}
        \frac{1}{2}(-\lambda)^\phi + \frac{1}{2} & (-1 \le \lambda < 0) \\
        -\frac{1}{2}(\lambda)^\phi + \frac{1}{2} & (0 \le \lambda < 1).
    \end{cases} 
\end{equation}

\begin{figure}
    \centering
    \includegraphics[width=\linewidth]{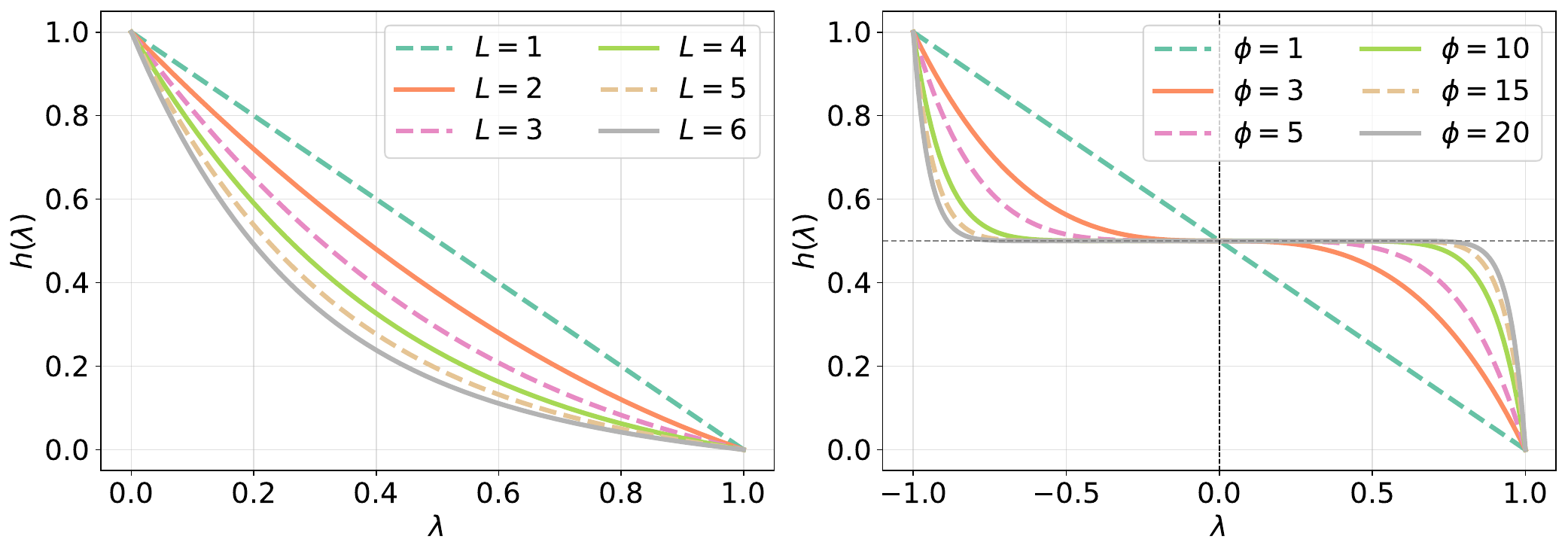}
    \vspace{-20pt}
    \caption{Visualization of the linear (left) and plateau (right) transfer functions. (Frequency $\lambda$ is rescaled to $[-1,1]$ for plateau as described in \cref{eq:cheby_filter}.)} 
    \label{fig:filter}
    \vspace{-0.4cm}
\end{figure}

As the number of layers $L$ in GCNs increases, the weights in the corresponding $L$-layered linear low-pass filter decay more rapidly, thereby narrowing the range of relatively emphasized low frequencies, as illustrated in \cref{fig:filter} (left).
This results in a simplistic filtering behavior that divides the spectrum into informative low and uninformative high frequencies, placing greater emphasis on the former, which typically encode global preference patterns.
In the plateau transfer function, on the other hand, \cref{fig:filter} (right) illustrates that the weights similarly drop at low frequencies, but unlike the linear case, they remain constantly over the mid-frequency range without significant decay.
This enables the filter to incorporate mid-frequency signals that capture fine-grained local preferences, though not as strongly emphasized as the low frequencies with global preferences.
Finally, the weights drop again at high frequencies, effectively suppressing strong high-frequency signals as noise.
In other words, the plateau function allows for more nuanced filtering to distinguish essential low-frequency, potentially helpful mid-frequency, and mostly noisy high-frequency signals, where the width of the mid-frequency region can be adjusted via the flatness parameter $\phi$.
Additionally, the function is smooth and symmetric, which facilitates its approximation.

\subsection{Filter Enhancement}
\label{sec:method:enhance}

We propose two modules, an ideal pass filter and degree-based normalization, to further enhance our Chebyshev filter.

\vspace{0.1cm} \noindent
\textbf{Ideal Pass Filter.}
The Chebyshev filter uses smooth polynomial bases, making it difficult to approximate step-like transfer functions with abrupt discontinuities unless the order is significantly increased, which comes at the cost of higher computation. Recall that CF tasks typically require emphasizing low-frequency signals over high-frequency ones. If this emphasis involves a sharp transition around a specific threshold, step-like behavior is needed, and Chebyshev filters alone may not be sufficient.

To address such scenarios, we introduce an ideal pass filter to enforce a sharp separation at a desired frequency threshold.
The transfer function for the ideal pass filter is $h_{\text{Ideal}}(\lambda)=\mathbb{I}[\lambda \le \lambda_\eta]$, where $\eta$ specifies the frequency index corresponding to the threshold above which the weights sharply decline.
The corresponding ideal pass filter is given by
\begin{equation}
    \label{eq:ideal_filter}
    H_{\text{Ideal}}(\mathbf{L}^\ast) = \mathbf{Q}^\ast \text{diag}(\underbrace{1, \dots, 1,}_{\eta} \underbrace{0, \dots, 0}_{|\mathcal{I}|-\eta}) {\mathbf{Q}^\ast}^\top \overset{(a)}{=} \mathbf{V}_\eta\mathbf{V}_\eta^\top,
\end{equation}
where $\mathbf{Q}^\ast$ represents the eigenvectors of $\mathbf{L}^\ast$, and $\mathbf{V}_\eta$ is the truncated right singular vectors of $\Tilde{\mathbf{R}}$ corresponding to $\eta$ largest singular values.
Here, Eq.~\eqref{eq:ideal_filter}-(a) is guaranteed by \cref{lem:b} in \cref{sec:proof}.
\cref{fig:method} illustrates that the comprehensive form gets a steep shape at $\lambda_\eta$, as the ideal pass filter (depicted in orange) is added to the Chebyshev filter (depicted in blue). 

Since the ideal pass filter requires singular value decomposition, it cannot leverage the computational advantage of matrix polynomials enjoyed by the Chebyshev filter.
However, it only requires a small number ($\eta$) of right singular vectors, eliminating the need for full decomposition.
This partial decomposition can be done efficiently using the generalized power method \cite{journeeGeneralizedPowerMethod2010}.

\vspace{0.1cm} \noindent
\textbf{Degree-based Normalization.}
Recall that a user's graph signal $\mathbf{r}_u \in \{0,1\}^{|\mathcal{I}|}$ is a binary vector indicating her interactions on items.
In order to prevent the signal from being over-dominated by popular items, we introduce a degree-based normalization.
In \cref{fig:method}, this normalization (depicted in yellow) is applied to the signal both before and after filtering.
Given a Chebyshev filter $H_\text{Cheby}$, it can be interpreted as an extended choice of graph Laplacian in the filter to a generalized (asymmetric) one by
\begin{equation}
    H_{\text{Cheby}}(\bar{\mathbf{L}}) = 
    H_{\text{Cheby}}(\mathbf{D}_\mathcal{I}^\beta \mathbf{L^\ast} \mathbf{D}_\mathcal{I}^{-\beta}) = \mathbf{D}_\mathcal{I}^\beta 
    H_{\text{Cheby}}(\mathbf{L^\ast}) \mathbf{D}_\mathcal{I}^{-\beta},
    \label{eq:normalization}
\end{equation}
where $\bar{\mathbf{L}} = \mathbf{D}_\mathcal{I}^\beta \mathbf{L^\ast} \mathbf{D}_\mathcal{I}^{-\beta}$ is the generalized graph Laplacian, $\mathbf{D}_\mathcal{I}$ is the degree matrix of the items defined in \cref{sec:prelim:gcn_for_recs}, and $\beta \geq 0$ is a power to control the strength of normalization.

Specifically, the dissimilarity between co-consumed items in the generalized graph Laplacian can be computed by
\begin{equation}
    (\bar{\mathbf{L}} \mathbf{r}_u)_i = \sum_{j \in \mathcal{N}(i)} w_{ij} \left( \frac{(\mathbf{r}_u)_i}{d_i} - \frac{(\mathbf{r}_u)_j}{d_i^{0.5-\beta} d_j^{0.5+\beta}} \right),
\end{equation}
where $\mathcal{N}(i)$ is the set of items co-consumed with item $i$, $d_i$ is the degree of item $i$, and $w_{ij} = \mathbf{r}_i^{\prime \top} \mathbf{D}_\mathcal{U}^{-1} \mathbf{r}_j^{\prime}$ is the edge weight derived from item-item co-occurrence.
$\mathbf{r}_i^{\prime}$ is the $i$-th column of $\mathbf{R}$.
As shown here, with a larger $\beta$, the signal value of popular items shrinks more.

\subsection{The Proposed Method: ChebyCF}
\label{sec:method:cheby_cf}

The overall structure of ChebyCF is illustrated in \cref{fig:method}.
Integrating the previously defined Chebyshev Filter (\cref{sec:method:cheby_filter}), Ideal Pass Filter and Degree-based Normalization (\cref{sec:method:enhance}) into the context of our graph (\cref{sec:method:graph}), we define the graph spectral filtering of our ChebyCF by
\begin{equation}
  \label{eq:overall}
  \hat{\mathbf{r}}_u = \mathbf{D}_\mathcal{I}^{\beta} (H_{\text{Cheby}}(\mathbf{L}^\ast;\phi) + \alpha H_{\text{Ideal}}(\mathbf{L}^\ast;\eta)) \mathbf{D}_\mathcal{I}^{-\beta} \mathbf{r}_u,
\end{equation}
where $\hat{\mathbf{r}}_u \in \mathbb{R}^{|\mathcal{I}|}$ is the predicted preference of user $u$ on all items, $\phi$ is the flatness of the plateau function, which is set as the target transfer function of $H_{\text{Cheby}}$, $\eta$ and $\alpha$ denote the weight of the ideal pass filter and the threshold frequency, respectively, while $\beta$ indicates the power of normalization.

The $K$-th order Chebyshev filter has linear computational complexity with respect to $K$, as the bases can be computed using a recurrence relation, which is defined in \cref{def:chebypoly}.
Consequently, similarly to other studies based on Chebyshev polynomials \cite{defferrardConvolutionalNeuralNetworks2016, heConvolutionalNeuralNetworks2022}, ChebyCF achieves efficient time complexity of $O(K\cdot\mathrm{nnz}(\mathbf{R}))$, where $\mathrm{nnz}(\mathbf{R})$ denotes the number of interactions.
\section{Experiments}
\label{sec:exp}

\subsection{Experimental Setup}
\label{sec:exp:setting}

\begin{table}[t]
    \centering
    \caption{Statistics of the datasets.}
    \vspace{-0.3cm}
    \scalebox{0.85}{
    \begin{tabular}{l|c|c|c|c}
    \Xhline{3.5\arrayrulewidth}
    \multicolumn{1}{c|}{\textbf{Dataset}}&  \# \textbf{User}&  \# \textbf{Item}&  \# \textbf{Interactions}&  \textbf{Density}\\
            \hline
                LastFM&  23,566&  48,123&  1,542,856 &  0.136\% \\ 
                Gowalla&  29,858&  40,981&  1,027,370&  0.084\% \\ 
                Amazon-book&  52,643&  91,599&  2,984,108&  0.062\% \\   
            \Xhline{3.5\arrayrulewidth}
    \end{tabular}}
    \label{tab:dataset}
    \vspace{-0.3cm}
\end{table}

\textbf{Datasets and Baselines}
We evaluate on three popular top-$N$ recommendation benchmark datasets summarized in \cref{tab:dataset}: LastFM \cite{wangKGAT2019}, Gowalla \cite{liangGowalla2016}, and Amazon-Book \cite{he2016ups}.
We compare with diverse state-of-the-art CF models, including classic non-graph models (MF-BPR \cite{korenMatrixFactorizationTechniques2009}, NeuMF \cite{heNeuralCollaborativeFiltering2017}, Mult-VAE \cite{liangVariationalAutoencodersCollaborative2018}), GCN-based models (NGCF \cite{wangNeuralGraphCollaborative2019}, LightGCN \cite{heLightGCNSimplifyingPowering2020}, UltraGCN \cite{maoUltraGCNUltraSimplification2021}, GTN \cite{fanGraphTrendFiltering2022}, SGCF \cite{heSGCF2023}, JGCF \cite{guoJGCF2023}, CAGCN(*-jc) \cite{wangCollaborationAwareGraphConvolutional2023}, Adap-$\tau$ \cite{chenAdapTau2023}, GTE \cite{caiGTE2023}, AFD-GCF \cite{wuAFDGCF2024}, GiffCF \cite{zhuGiffCF2024}), and GSF-based models (GF-CF \cite{shenHowPowerfulGraph2021}, LinkProp(-Multi) \cite{fuRevisitingNeighborhoodbasedLink2022}, PGSP \cite{liuPersonalizedGraphSignal2023}, BSPM \cite{choiBSPM2023}, SGFCF \cite{pengSGFCF2024}).
We use the LightGCN backbone for Adap-$\tau$ \cite{chenAdapTau2023} and AFD-GCF \cite{wuAFDGCF2024}.

\vspace{0.1cm} \noindent
\textbf{Evaluation Protocol.}
For a fair comparison, we use the same train/test splits and metrics as previous studies \cite{wangNeuralGraphCollaborative2019, heLightGCNSimplifyingPowering2020, maoUltraGCNUltraSimplification2021, shenHowPowerfulGraph2021, fanGraphTrendFiltering2022, liuPersonalizedGraphSignal2023, choiBSPM2023}. 
The metrics are Recall and Normalized Discounted Cumulative Gain (NDCG), which evaluate how many items from the test set are included in the top-$N$ ranked list of items according to the user's predicted preference. 
Being consistent with prior studies, we report results with $N=20$ by default and additionally include results with $N=10$ in overall performance.

\vspace{0.1cm} \noindent
\textbf{Implementation Details.}
For the baselines, we follow the training configurations including hyperparameters as described in the original papers or official codes.
For our ChebyCF, we set the order ($K$) as 8, grid-search the flatness of plateau ($\phi$) from 1 to 20 in the step size of 0.5, the weight ($\alpha$) from 0.0 to 0.5 with step size 0.1, the threshold frequency ($\eta$) in $\{2^7, 2^8, 2^9, 2^{10}, 2^{11}\}$, and the power ($\beta$) from 0.0 to 0.5 in the step size of 0.1.

\begin{figure}
    \centering
    \includegraphics[width=\linewidth]{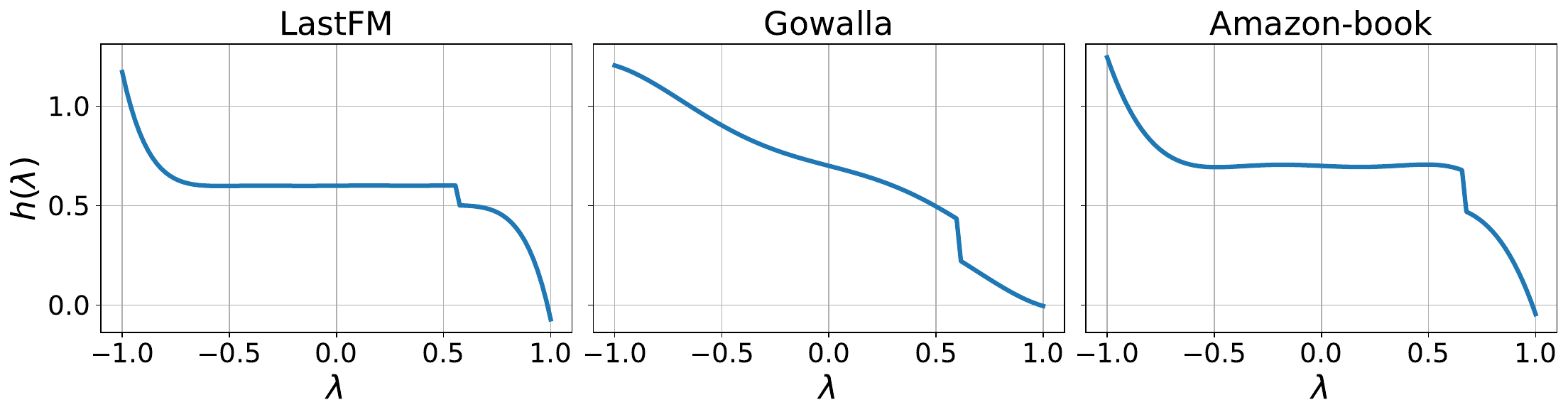}
    \vspace{-22pt}
    \caption{The best transfer functions found for ChebyCF.}
    \label{fig:best_filter}
    \vspace{-0.2cm}
\end{figure}

\subsection{Overall Performance}
\begin{table*}
\caption{Overall performance comparison. The highest performance is bold-faced, and the second best is underlined. (Reported scores brought from the original papers are marked with $^*$.)}
\vspace{-0.3cm}
\centering
\scalebox{.90}{
\begin{tabular}{l|l|cccc|cccc|cccc}
\Xhline{3.5\arrayrulewidth}
\multicolumn{1}{c|}{\multirow{3}{*}{\textbf{Category}}} & \multicolumn{1}{c|}{\multirow{3}{*}{\textbf{Model}}}  & \multicolumn{4}{c|}{\textbf{LastFM}} & \multicolumn{4}{c|}{\textbf{Gowalla}} & \multicolumn{4}{c}{\textbf{Amazon-book}} \\
    &                 & \multicolumn{2}{c}{Recall@\textit{N}} & \multicolumn{2}{c|}{NDCG@\textit{N}} & \multicolumn{2}{c}{Recall@\textit{N}} & \multicolumn{2}{c|}{NDCG@\textit{N}} & \multicolumn{2}{c}{Recall@\textit{N}} & \multicolumn{2}{c}{NDCG@\textit{N}} \\
    &                 & ${N=10}$ & ${N=20}$ & ${N=10}$ & ${N=20}$ & ${N=10}$ & ${N=20}$ & ${N=10}$ & ${N=20}$ & ${N=10}$ & ${N=20}$ & ${N=10}$ & ${N=20}$ \\
\Xhline{2.5\arrayrulewidth}
\multirow{3}{*}{Non-graph}
    & MF-BPR          & 0.0505        & 0.0743        & 0.0558        & 0.0627        & 0.0958        & 0.1387        & 0.1032        & 0.1158        & 0.0175        & 0.0312        & 0.0188        & 0.0242 \\
    & NeuMF           & 0.0515        & 0.0727        & 0.0559        & 0.0621        & 0.0925        & 0.1344        & 0.1000        & 0.1121        & 0.0158        & 0.0276        & 0.0169        & 0.0216 \\
    & Mult-VAE        & 0.0545        & 0.0747        & 0.0622        & 0.0672        & 0.1088        & 0.1567        & 0.1152        & 0.1297        & 0.0234        & 0.0403        & 0.0242        & 0.0310 \\
\hline
\multirow{11}{*}{GCN-based}
    & NGCF            & 0.0542        & 0.0783        & 0.0609        & 0.0677        & 0.1153        & 0.1636        & 0.1256        & 0.1392        & 0.0197        & 0.0349        & 0.0207        & 0.0267 \\
    & LightGCN        & 0.0582        & 0.0832        & 0.0672        & 0.0738        & 0.1257        & 0.1786        & 0.1381        & 0.1526        & 0.0238        & 0.0413        & 0.0253        & 0.0322 \\
    & UltraGCN        & 0.0609        & 0.0867        & 0.0706        & 0.0772        & 0.1301        & 0.1849        & 0.1409        & 0.1565        & 0.0424        & 0.0683        & 0.0459        & 0.0558 \\
    & GTN             & 0.0659        & 0.0936        & 0.0795        & 0.0863        & 0.1324        & 0.1884        & 0.1427        & 0.1590        & 0.0248        & 0.0433        & 0.0259        & 0.0333 \\
    & SGCF*           &  -            & -            & -              &  -            & -            & 0.1862         &  -            & 0.1580       & -              & 0.0466        &  -   & 0.0358 \\
    & JGCF            & 0.0621        & 0.0864        & 0.0728        & 0.0786        & 0.1299        & 0.1837        & 0.1403        & 0.1553        & 0.0296        & 0.0495        & 0.0312        & 0.0390 \\
    & CAGCN(*-jc)     & 0.0537        & 0.0727        & 0.0678        & 0.0712        & 0.1308        & 0.1878        & 0.1432        & 0.1591        & 0.0308        & 0.0510        & 0.0325        & 0.0403 \\
    & Adap-$\tau$     & 0.0673        & 0.0942        & 0.0779        & 0.0846        & 0.1327        & 0.1901        & 0.1426        & 0.1590        & 0.0375        & 0.0612        & 0.0399        & 0.0490 \\
    & GTE             & 0.0645        & 0.0872        & 0.0742        & 0.0793        & 0.0963        & 0.1369        & 0.1026        & 0.1142        & 0.0211        & 0.0348        & 0.0224        & 0.0278 \\
    & AFD-GCF         & 0.0678        & 0.0951        & 0.0789        & 0.0858        & 0.1284        & 0.1820        & 0.1407        & 0.1556        & 0.0263        & 0.0458        & 0.0277        & 0.0354 \\
    & GiffCF          & \scnd{0.0885} & 0.1025        & 0.0894        & 0.0946        & 0.1125        & 0.1478        & 0.1138        & 0.1271        & \best{0.0474} & 0.0670        & 0.0472        & 0.0566 \\
\hline
\multirow{6}{*}{GSF-based}                                
    & GF-CF           & 0.0812        & 0.1107        & 0.0937        & 0.1011        & 0.1279        & 0.1850        & 0.1350        & 0.1518        & 0.0450        & 0.0710        & 0.0484        & 0.0584 \\
    & LinkProp(-Multi)*&  -            & 0.1071        &  -            & 0.1039        &  -            & 0.1908        &  -            & 0.1573        &  -            & 0.0721        &  -            & 0.0588 \\
    & PGSP            & 0.0827        & 0.1136        & 0.0962        & 0.1037        & 0.1344        & 0.1917        & \scnd{0.1443} & \scnd{0.1606} & 0.0448        & 0.0710        & 0.0483        & 0.0584 \\
    & BSPM        & 0.0860        & \scnd{0.1157} & \scnd{0.0984} & \scnd{0.1057} & \scnd{0.1345} & \scnd{0.1921} & 0.1432        & 0.1597        & \scnd{0.0473}        & \scnd{0.0733} & \scnd{0.0511} & \scnd{0.0610} \\
    & SGFCF           & 0.0842        & 0.1143        & \scnd{0.0984} & 0.1054        & 0.1324        & 0.1898        & 0.1398        & 0.1566        & 0.0392        & 0.0640        & 0.0423        & 0.0519 \\ 
\hhline{~-------------}
    & \textbf{ChebyCF (Ours)}  & \best{0.0892} & \best{0.1199} & \best{0.1027} & \best{0.1101} & \best{0.1357} & \best{0.1941} & \best{0.1448} & \best{0.1616} & \best{0.0474} & \best{0.0738} & \best{0.0513} & \best{0.0613} \\
\Xhline{3.5\arrayrulewidth}
\end{tabular}
}
\label{tab:performance}
\end{table*}

We compare the overall performance of ChebyCF and the baselines in \cref{tab:performance}, and illustrate the best transfer functions found for ChebyCF in \cref{fig:best_filter}.
We observe that ChebyCF consistently achieves the best performance across all datasets.
Given that all GSF baselines directly apply spectral filtering without embedding layers, where GF-CF, PGSP, and SGFCF particularly emphasizes low frequencies through ideal pass filters just like ChebyCF, the main difference to make ChebyCF outperform them is its more generalized filter structure, which captures and utilizes mid-frequency components potentially beneficial for personalization, unlike the linear low-pass filter forms employed by the baselines.
GSF-based models, which use the entire user interaction as input without an embedding layer, achieve higher performance than GCN-based models with an embedding layer, suggesting the benefit of full exploitation of the user behavior.
GCN-based models consistently outperform non-graph models, likely due to their ability to capture collaborative signals within high-order connectivity.

\subsection{Effectiveness Analysis}
\label{sec:exp:ablation}

We analyze the effectiveness of ChebyCF based on the two hypotheses proposed in \cref{sec:motiv}:
1) accommodating the full spectrum of frequency information, including high frequencies, would enable superior personalized recommendations when appropriately filtered, and 2) employing a more sophisticated filtering beyond simple linear low-pass filtering would lead to more refined recommendations.

\begin{figure}
    \centering
    \includegraphics[width=\linewidth]{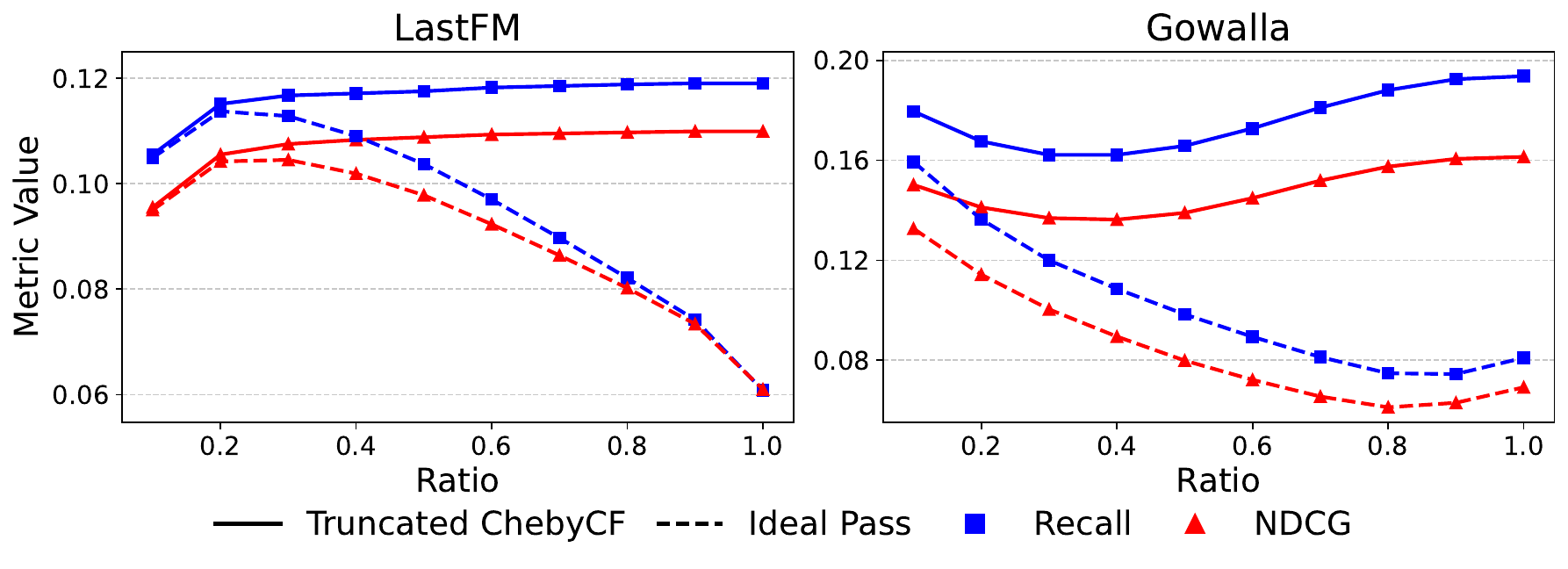}
    \vspace{-20pt}
    \caption{Performance comparison when utilizing a specific ratio of lower frequencies.}
    \label{fig:truncated}
    \vspace{-0.3cm}
\end{figure}

\vspace{0.1cm} \noindent
\textbf{Effect of Full-range Frequencies.}
\label{sec:exp:full-range}
To validate the first hypothesis, we compare the full ChebyCF against itself with a restricted range of frequencies.
Specifically, a truncated version of ChebyCF is designed as a filter whose transfer function is identical to ChebyCF for the lower frequencies corresponding to a specified ratio, while it is set to 0 for the remaining high frequencies. Therefore, when the ratio is 1.0, it is identical to the original ChebyCF.
Additionally, we also compare with the ideal pass filter with the same frequency ranges, to verify the effect of our enhanced filter.
For each method, we vary the ratio from 0.1 to 1.0.
As this experiment involves filtering with a non-polynomial transfer function that truncates at a specific frequency, the full SVD of $\Tilde{\mathbf{R}}$ is needed.
Due to its computational overhead, we perform this experiment only on LastFM and Gowalla.

\cref{fig:truncated} presents the results.
First, we observe that the performance of our ChebyCF (solid lines in \cref{fig:truncated}) monotonically improves with more high-frequency components accommodated (that is, with a larger `ratio').
This result concludes that the high-frequency components indeed contain signals useful for personalized recommendations.
Comparing it with the curve with the ideal pass filter (dotted lines in \cref{fig:truncated}), however, reveals that we have not effectively utilized this high-frequency components with the previous methods.
As seen in the figure, the ideal pass filter achieves its best performance when the filter retains only the lowest 10-20\% frequencies and shows a significant performance drop thereafter.
This observation highlights the valuable information embedded in higher frequencies cannot be effectively utilized without precise filtering.
To conclude, high-frequency components indeed contain valuable information essential for personalized recommendations, but this information can be effectively refined and utilized only by employing a well-designed filter like ChebyCF.

\begin{table}
\centering
\caption{Comparison with $L$-layered linear low-pass filters.}
\vspace{-0.3cm}
\scalebox{0.85}{
\begin{tabular}{l|cc|cc|cc}
\Xhline{3.5\arrayrulewidth} & \multicolumn{2}{c|}{\textbf{LastFM}} & \multicolumn{2}{c|}{\textbf{Gowalla}} & \multicolumn{2}{c}{\textbf{Amazon-book}} \\
                & Recall        & NDCG          & Recall        & NDCG           & Recall        & NDCG           \\ 
\hline
\textbf{ChebyCF (Ours)}  & \best{0.1199} & \best{0.1101} & \best{0.1941} & \best{0.1616}  & \best{0.0738} & \best{0.0613}  \\ 
LightGCN        & 0.0832        & 0.0738        & 0.1786        & 0.1526         & 0.0413        & 0.0322         \\

\hline
1-layer         & 0.1107          & 0.1011      & 0.1682        & 0.1331         & 0.0710        & 0.0584 \\
2-layer         & 0.1099          & 0.1002      & 0.1736        & 0.1389         & 0.0683        & 0.0560 \\
3-layer         & 0.1098          & 0.0999      & 0.1746        & 0.1403         & 0.0659        & 0.0538 \\
4-layer         & 0.1096          & 0.0994      & 0.1737        & 0.1403         & 0.0637        & 0.0519 \\
\Xhline{3.5\arrayrulewidth}         
\end{tabular}
}
\label{tab:linear}
\end{table}

\vspace{0.1cm} \noindent
\textbf{Effect of the Flexible Filtering.}
We compare ChebyCF with a linear low-pass filter, which corresponds to the neighborhood aggregation as demonstrated in Theorem \ref{theorem:bottleneck}.
For comprehensive comparison, we include $L$-layered linear low-pass filters, corresponding to $L$-layered GCNs that aggregate up to $L$-hop neighbors.
Hence, the transfer functions are generalized from $(1 - \lambda)$ to $\frac{1}{L} \sum_{l=1}^L (1 - \lambda)^l$.
To isolate the impact of the filter function, we exclude the embedding layer from all methods, preventing any cutoff in the filter.

\cref{tab:linear} compares the performance of ChebyCF against LightGCN~\cite{heLightGCNSimplifyingPowering2020}, a representative baseline model employing an embedding layer with linear low-pass filtering, and generalized linear low-pass filters with $L$ stacked layers.
We observe that ChebyCF significantly outperforms all other baselines, indicating the critical role of elaborate filter functions in capturing and leveraging frequency-specific information for the recommendation task.

Taking a deeper look, we interpret the effect of the plateau function used as the transfer function in ChebyCF.
As shown on the right side of \cref{fig:filter}, it narrows the range of the focused low frequencies as $\phi$ increases, while also incorporates information from mid-frequencies.
In contrast, as shown on the left side of \cref{fig:filter}, the linear low-pass filter can narrow the range of the focused lower frequencies as more layers are stacked, but cannot be modified to exploit higher frequencies.
This fundamental limitation illustrates the underlying cause of the over-smoothing issue \cite{liuInterestawareMessagePassingGCN2021, wuAFDGCF2024} in GCN-based CF models, where stacking more layers to expand the receptive field results in a performance drop.
A similar performance drop with stacked linear low-pass filters is observed in \cref{tab:linear}.

Additionally, comparing the 1-layer and LightGCN, where both of them adopt the same linear low-pass filter but only LightGCN incorporates an embedding layer, we observe that LightGCN significantly underperforms on LastFM and Amazon-book.
This also aligns with our earlier hypothesis that embedding layers may hinder the effective utilization of full frequency information.

\subsection{Ablation Studies}

\begin{figure}
    \centering
    \includegraphics[width=\linewidth]{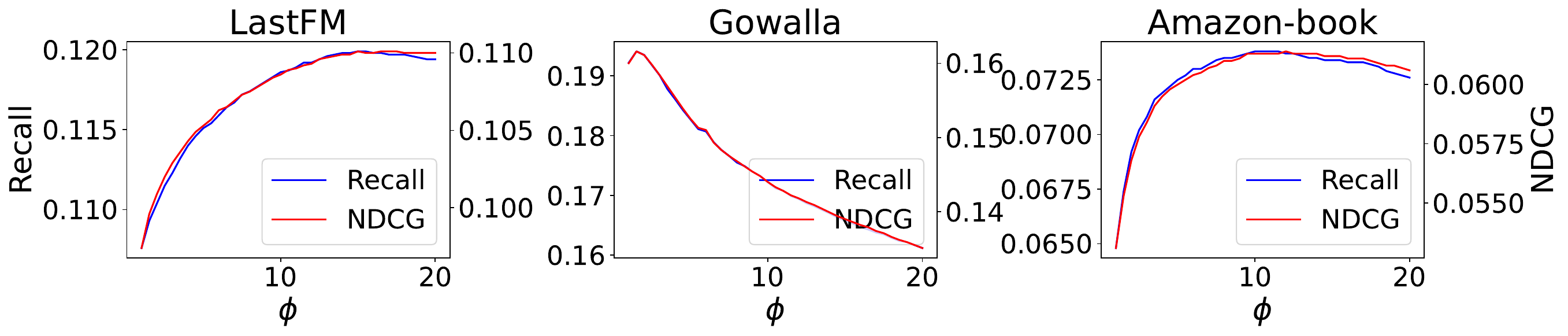}
    \vspace{-20pt}
    \caption{Performance by varying $\phi$.}
    \label{fig:phi}
    \vspace{-0.4cm}
\end{figure}

\begin{figure}
    \centering
    \includegraphics[width=\linewidth]{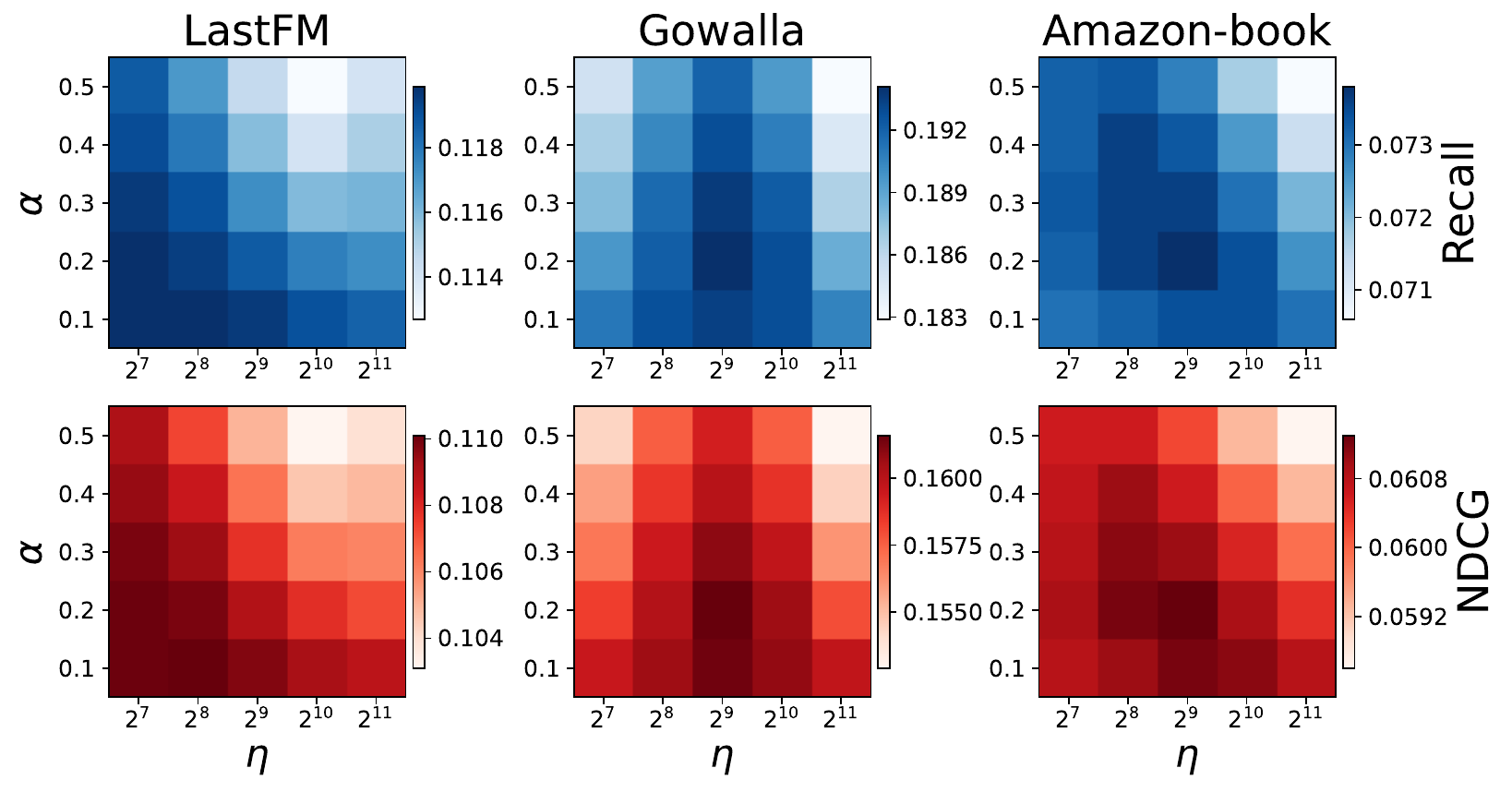}
    \vspace{-20pt}
    \caption{Performance by varying $\alpha$ and $\eta$.}
    \label{fig:ideal}
    \vspace{-0.4cm}
\end{figure}

\begin{figure}
    \centering
    \includegraphics[width=\linewidth]{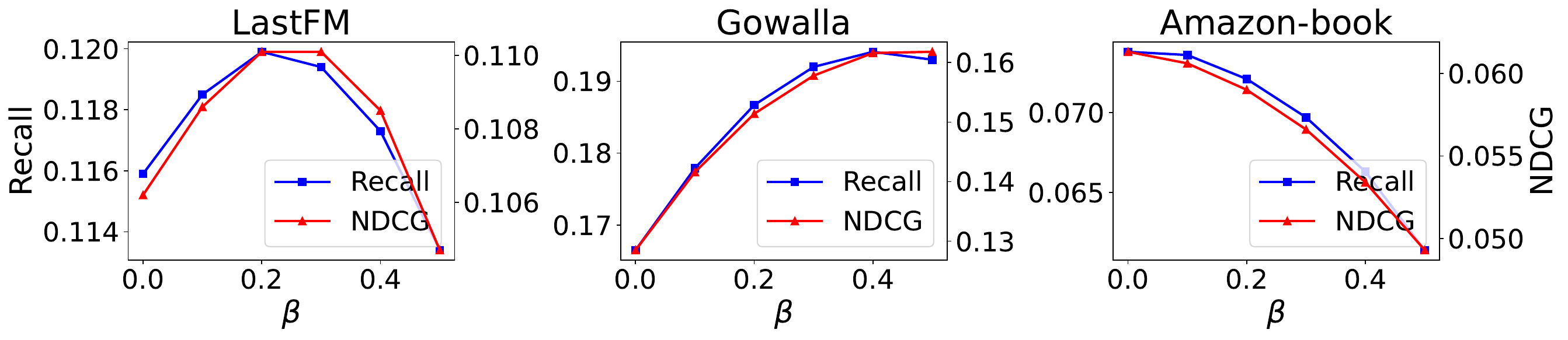}
    \vspace{-20pt}
    \caption{Performance by varying $\beta$.}
    \label{fig:norm}
    \vspace{-0.2cm}
\end{figure}

To analyze the role of each module in ChebyCF, we conduct ablation studies and demonstrate the effect of hyper-parameters.
\cref{tab:ablation} reports the performance of ChebyCF when each module is excluded, and \cref{fig:phi}-\ref{fig:norm} illustrate the performance variations with varied hyperparameters of the Chebyshev filter ($\phi$ in Eq.~\eqref{eq:plateau}), ideal pass filter ($\eta$ in Eq.~\eqref{eq:ideal_filter} and $\alpha$ in Eq.~\eqref{eq:overall}), and degree-based normalization ($\beta$ in Eq.~\eqref{eq:normalization}), respectively.

First, using a simple linear filter instead of the Chebyshev filter always drops the performance.
This highlights the advantage of the Chebyshev interpolation, which allows greater flexibility in adjusting the filter's form.
Interestingly, the performance drop is relatively smaller on Gowalla.
Observing its optimal filter shown in \cref{fig:best_filter} closely resembles a linear filter, we conclude that this is because a linear filter is sufficient for this dataset.
This is also aligned with \cref{fig:phi}. On Gowalla, the performance drops as $\phi$ deviates from 1, where the plateau function gets closer to linear, as shown in \cref{fig:filter}.

Besides the Chebyshev Filter, the degree-based normalization drops the performance most on LastFM and Gowalla when removed.
In contrast, Amazon-Book shows no performance difference, achieving the best performance when the normalization is absent ($\beta = 0$), according to \cref{fig:norm}.
This indicates that the degree of popularity bias varies across datasets, and the normalization should be adjusted accordingly to account for these differences.

Lastly, \cref{fig:ideal} shows that the model performs optimally with a moderate choice of $\alpha$ and $\eta$.
This confirms that the role of the ideal pass filter, allowing the weight for the highest frequencies to steeply drop to filter out signals closer to noise, works as intended.

In conclusion, ChebyCF achieves superior performance by centering on the Chebyshev filter, which enables highly nuanced and flexible filtering tailored to the characteristics of each dataset.
This is complemented by the ideal pass filter, which handles steep changes in importance across frequency ranges, and the degree-based normalization, which addresses severe popularity bias. Together, these components form a robust and effective framework.

\begin{table}
\centering
\caption{Performance with different ablation settings.}
\vspace{-0.3cm}
\scalebox{0.8}{
\begin{tabular}{r|cc|cc|cc}
\Xhline{3.5\arrayrulewidth} 
\multicolumn{1}{c|} {\multirow{2}{*}{\textbf{Model}}} & \multicolumn{2}{c|}{\textbf{LastFM}} & \multicolumn{2}{c|}{\textbf{Gowalla}} & \multicolumn{2}{c}{\textbf{Amazon-book}} \\
                & Recall      & NDCG     & Recall      & NDCG      & Recall     & NDCG          \\ 
\Xhline{2.5\arrayrulewidth} 
\multicolumn{1}{c|}{\textbf{ChebyCF (Ours)}}  & \best{0.1199}	  & \best{0.1101}	 & \best{0.1941}	   & \best{0.1616}	   & \best{0.0738}	    & \best{0.0613} \\
\hline
w/o Chebyshev Filter   & 0.1107	& 0.1011	& 0.1928	& 0.1606	& 0.0675	& 0.0554 \\
w/o Normalization      & 0.1159	& 0.1062	& 0.1665	& 0.1285	& 0.0738	& 0.0613 \\
w/o Ideal Pass Filter  & 0.1194	& 0.1096	& 0.1915	& 0.1601	& 0.0725	& 0.0606 \\
\Xhline{3.5\arrayrulewidth}         
\end{tabular}
}
\vspace{-0.4cm}
\label{tab:ablation}
\end{table}

\subsection{Computational Efficiency}
\label{sec:exp:computation}

We compare the computational efficiency at inference between our ChebyCF and several baselines: MF-BPR \cite{rendleBPRBayesianPersonalized2009}, LightGCN \cite{heLightGCNSimplifyingPowering2020}, GF-CF \cite{shenHowPowerfulGraph2021}, and BSPM \cite{choiBSPM2023}, the current state-of-the-art in CF recommendations inspired by score-based generative models \cite{vahdatSGMs2021}.

\cref{tab:Efficiency} reports the per-user inference time and the corresponding Recall metrics of competing methods.
Compared with BSPM, the top-performing method, our method takes significantly less inference time (up to 8.1 times faster) while achieving comparable or slightly superior Recall.
Even with a more flexible filter, our ChebyCF performs inference reasonably fast compared to lighter models such as (LightGCN \cite{heLightGCNSimplifyingPowering2020}, MF-BPR \cite{rendleBPRBayesianPersonalized2009}, and GF-CF \cite{shenHowPowerfulGraph2021}), thanks to the enhanced computational efficiency achieved by the Chebyshev filter, a graph filter that is free from matrix decomposition, as mentioned in \cref{sec:method:cheby_filter}.

\begin{table}
\centering
\caption{Comparison of per-example inference time.}
\vspace{-0.3cm}
\scalebox{0.82}{
\begin{tabular}{l|cc|cc|cc}
\Xhline{3.5\arrayrulewidth} 
\multicolumn{1}{c|} {\multirow{2}{*}{\textbf{Model}}} & \multicolumn{2}{c|}{\textbf{LastFM}} & \multicolumn{2}{c|}{\textbf{Gowalla}} & \multicolumn{2}{c}{\textbf{Amazon-book}} \\
                                & Time        & Recall   & Time        & Recall    & Time        & Recall          \\ 
\Xhline{2.5\arrayrulewidth} 
MF-BPR          & 0.303ms     & 0.0743   & 0.269ms     & 0.1387    & 0.549ms    & 0.0312    \\
LightGCN        & 0.304ms     & 0.0832   & 0.263ms     & 0.1786    & 0.578ms    & 0.0413    \\
GF-CF           & 0.120ms     & 0.1107   & 0.107ms     & 0.1850    & 0.226ms    & 0.0710    \\
BSPM-Euler      & 0.545ms     & 0.1149   & 0.319ms     & 0.1864    & 1.465ms    & 0.0729    \\
BSPM-RK4        & 1.789ms     & 0.1157   & 0.751ms     & 0.1921    & 6.781ms    & 0.0733    \\
\hline
\textbf{ChebyCF (Ours)} & 0.391ms & 0.1199   & 0.293ms      & 0.1941    & 0.834ms     & 0.0738    \\
\Xhline{3.5\arrayrulewidth}         
\end{tabular}
}
\label{tab:Efficiency}
\vspace{-0.4cm}
\end{table}
\section{Related Works}
\label{sec:related}

\textbf{Graph Neural Networks.}
Beside random-walk based approaches \cite{perozziDeepwalk2014, groverNode2vecScalableFeature2016, huangGraphRecurrentNetworks2019, nikolentzosRandomwalkgraphneuralnetworks2020, jinRawgnn2022, wangNonConvGNN2024}, there are two primary approaches in Graph Neural Networks (GNNs): spatial and spectral
graph convolutions.
Spatial graph convolution defines convolution in the vertex domain.
GCN \cite{kipfSemiSupervisedClassificationGraph2017} simplifies graph spectral convolutions using only the first-order linear filters, equivalent to the spatial convolution of 1-hop neighbor aggregation.
SGC \cite{wuSimplifyingGraphConvolutional2019} reduces its computational complexity by removing its redundant feature transformations.
More recent works have expanded GCNs to a broader range of applications \cite{hamiltonInductiveRepresentationLearning2017, gilmerNeuralMessagePassing2017, velickovicDeepGraphInfomax2018, xuHowPowerfulAre2019}.

Spectral graph convolution relies on costly Laplacian eigendecomposition to perform convolution in the spectral domain, triggering the development of numerous polynomial approximations to circumvent this issue.
ChebNet \cite{defferrardConvolutionalNeuralNetworks2016}, for example, adopts a Chebyshev basis.
GPR-GNN \cite{chienAdaptiveUniversalGeneralized2021} and BernNet \cite{heBernNetLearningArbitrary2021} utilize the monomial and Bernstein basis, respectively.
ChebNetII \cite{chenRevisitingGraphBased2020} presents an improved model using Chebyshev interpolation, which reduces the Runge phenomenon.
Inspired by the strong performance of graph filtering using polynomial approximations in the node classification task, we design ChebyCF—a graph filter utilizing Chebyshev interpolation, specifically adapted to the context of collaborative filtering, including the elimination of heavy node-wise feature transformations.

\vspace{0.1cm} \noindent
\textbf{Graph-based Recommendations.}
Interpreting the user-item interaction as a graph, GNNs have been extensively explored in collaborative filtering.
After GCNs are applied to CF for the first time \cite{wangNeuralGraphCollaborative2019}, several following studies \cite{chenRevisitingGraphBased2020, heLightGCNSimplifyingPowering2020} have shown that the inherent complexity of GCNs is less suitable for CF.
This is because CF relies only on the user-item interaction data without any feature information.
To address this, several attempts to simplify the model structure have been made \cite{maoUltraGCNUltraSimplification2021, heSGCF2023}.
There are also various approaches \cite{sunNeighborInteractionAware2020, wangDisentangledGraphCollaborative2020, liuInterestawareMessagePassingGCN2021, kongLinearNonLinearThat2022, fanGraphTrendFiltering2022, guoJGCF2023, wangCollaborationAwareGraphConvolutional2023, zhuGiffCF2024,jinri2024content,eungi2025reducedgcn} to capture important information from the user-item interaction graph.
Additionally, there have been attempts \cite{wuSelfsupervisedGraphLearning2021, xiaHypergraphContrastiveCollaborative2022, linImprovingGraphCollaborative2022, jiangAdaptiveGraphContrastive2023} to overcome the sparsity of interaction data by leveraging graph contrastive learning.

While many of the aforementioned studies have proposed GCN-based models with spatial convolution, there are also GSF-based approaches utilizing spectral convolution \cite{zhengSpectralCollaborativeFiltering2018, shenHowPowerfulGraph2021, fuRevisitingNeighborhoodbasedLink2022, liuPersonalizedGraphSignal2023, pengSGFCF2024, park2024turbo}.
LinkProp \cite{fuRevisitingNeighborhoodbasedLink2022} interprets CF from the perspective of the graph link prediction task.
GF-CF \cite{shenHowPowerfulGraph2021} highlights that the power of existing spatial convolution approaches lies in their low-pass filtering capabilities from the spectral filtering perspective.
PGSP \cite{liuPersonalizedGraphSignal2023} builds on GF-CF by augmenting both the input signal and graph.
SGFCF \cite{pengSGFCF2024} further enhances GF-CF through the introduction of a new graph normalization technique and individualized filtering.
Nevertheless, the form of graph filters has largely remained linear, with limited attention given to enabling more flexible formulations. Our method addresses this by leveraging Chebyshev interpolation with a graph filter that is both flexible and computationally efficient.
\section{Summary}
\label{sec:summary}

We present ChebyCF, a collaborative filtering framework grounded in graph spectral filtering (GSF). Unlike prior approaches that rely on embedding layers to project users and items into a constrained latent space, ChebyCF directly leverages raw interaction signals on the graph. It approximates a flexible non-linear graph filter via Chebyshev interpolation and further refines this process through the integration of ideal pass filters and degree-based normalization. Through extensive experiments, we show that ChebyCF effectively overcomes key bottlenecks of GCN-based collaborative filtering, achieving state-of-the-art performance across multiple recommendation benchmarks. We hope this work provides a natural transition from GCN-based models to GSF-based methods, and we envision future efforts focusing on developing principled supervision strategies for learning expressive spectral filters

{\subsubsection*{Acknowledgements}
This work was supported by Youlchon Foundation (Nongshim Corp.), NRF grants (RS-2021-NR05515, RS-2024-00336576, RS-2023-0022663) and IITP grants (RS-2022-II220264, RS-2024-00353131) by the government of Korea.}
\appendix

\renewcommand\thetable{\Roman{table}}
\renewcommand\thefigure{\Roman{figure}}
\setcounter{table}{0}
\setcounter{figure}{0}

\section{Appendix: Proof of Theorem 3.1}
\label{sec:proof}

\noindent \textit{Theorem \ref{theorem:bottleneck}.
Consider the \textit{Linear Graph Convolutional Network} (LGCN) \citep{heLightGCNSimplifyingPowering2020}, where propagation is performed via the neighborhood aggregation as in \cref{eq:gcn-cf}-(ii), and
the embedding matrices $\hat{\mathbf{E}}_\mathcal{U} \in \mathbb{R}^{\mathcal{|U|}  \times d }$ and $\hat{\mathbf{E}}_\mathcal{I} \in \mathbb{R}^{\mathcal{|I|}  \times d }$ are set by solving
\begin{align*}
  \left( \hat{\mathbf{E}}_\mathcal{U}, \hat{\mathbf{E}}_\mathcal{I} \right) &= \underset{(\mathbf{E}_\mathcal{U}, \mathbf{E}_\mathcal{I})}{\mathrm{argmin}}\ \| \Tilde{\mathbf{R}} - \mathbf{E}_\mathcal{U} \mathbf{E}_\mathcal{I}^\top \|_F 
  \ \  \text{s.t.}  \ \ 
  \mathrm{rank}(\mathbf{E}_\mathcal{U}),\mathrm{rank}(\mathbf{E}_\mathcal{I}) \le d, \end{align*}
with $d \le \min(|\mathcal{U}|, |\mathcal{I}|)$ as the embedding dimension and $\Tilde{\mathbf{R}}$ as the normalized interaction matrix.
Then, $\hat{\mathbf{r}}_u \in \mathbb{R}^{|\mathcal{I}|}$, the predicted preference of user $u$,
is equivalent to the spectrally filtered output of the input signal $\Tilde{\mathbf{r}}_u$ (\textit{i.e.}, the $u$-th row of $\Tilde{\mathbf{R}}$) over the normalized item-item graph with Laplacian $\mathbf{L}^\ast = \mathbf{I} - \Tilde{\mathbf{R}}^{\top}\Tilde{\mathbf{R}}$, using the transfer function $h:[0,1] \to [0,1]$ defined by $h(\lambda) = (1-\lambda) \cdot \mathbb{I}[\lambda \le \lambda_d]$, where $\mathbb{I}$ is indicator function and $\lambda_d$ is the $d$-th smallest frequency.}

\vspace{0.1cm}
\noindent \textit{Proof.}

Consider the normalized interaction matrix $\Tilde{\mathbf{R}} \in \mathbb{R}^{|\mathcal{U}| \times |\mathcal{I}|}$ and its singular value decomposition $\Tilde{\mathbf{R}} = \mathbf{U}\mathbf{\Sigma} \mathbf{V}^\top$, where the singular values are in descending order.
Let $\mathbf{U}_d \in \mathbb{R}^{|\mathcal{U}| \times d}$, $\mathbf{\Sigma}_d \in \mathbb{R}^{d \times d}$, and $\mathbf{V}_d \in \mathbb{R}^{|\mathcal{I}| \times d}$ be the left submatrix of $\mathbf{U}$, $\mathbf{\Sigma}$, and $\mathbf{V}$, respectively.
We define the partitioning matrix $\mathbf{P}_a = [\mathbf{I} \ \ \mathbf{O}]^\top \in \mathbb{R}^{a \times d}$ where $\mathbf{I}$ is an identity matrix and $\mathbf{O}$ is a zero matrix.
Considering the embedding propagated through one layer as the final embedding, $\mathbf{E}_{\mathcal{U}}^\ast$ and $\mathbf{E}_{\mathcal{I}}^\ast$ satisfy 
$\mathbf{E}^\ast_{\mathcal{U}} = \Tilde{\mathbf{R}} \hat{\mathbf{E}}_\mathcal{I}$ and $\mathbf{E}^\ast_{\mathcal{I}} = \Tilde{\mathbf{R}}^\top \hat{\mathbf{E}}_\mathcal{U}$, based on \cref{eq:gcn-cf}-(ii).
From $\hat{\mathbf{R}} = \mathbf{E}^\ast_{\mathcal{U}}{\mathbf{E}^\ast}_{\mathcal{I}}^\top$, we deduce $\hat{\mathbf{R}} = \Tilde{\mathbf{R}} \hat{\mathbf{E}}_\mathcal{I} \hat{\mathbf{E}}_\mathcal{U}^\top \Tilde{\mathbf{R}}$, and thus obtain the $u$-th row of $\hat{\mathbf{R}}$, $\hat{\mathbf{r}}_u = \Tilde{\mathbf{r}}_u (\hat{\mathbf{E}}_\mathcal{U}\hat{\mathbf{E}}_\mathcal{I}^\top)^\top \Tilde{\mathbf{R}}$.
From this,
\begin{alignat*}{3}
    \hat{\mathbf{r}}_u 
    & = \Tilde{\mathbf{r}}_u (\hat{\mathbf{E}}_\mathcal{U} \hat{\mathbf{E}}_\mathcal{I}^\top)^\top \Tilde{\mathbf{R}}
    = \Tilde{\mathbf{r}}_u  (\mathbf{U}_d \mathbf{\Sigma}_d \mathbf{V}_d^\top)^\top \Tilde{\mathbf{R}} \quad (\because \text{Lemma}\,\ref{lem:a}) \\
    &= \Tilde{\mathbf{r}}_u (\mathbf{U}\mathbf{P}_{|\mathcal{U}|} \mathbf{\Sigma}_d \mathbf{P}_{|\mathcal{I}|}^\top \mathbf{V}^\top)^\top \Tilde{\mathbf{R}}  \\
    &= \Tilde{\mathbf{r}}_u (\mathbf{V} \mathbf{\Sigma}^\top \mathbf{U}^\top \mathbf{U} \mathbf{P}_{|\mathcal{U}|} \mathbf{\Sigma}_d \mathbf{P}_{|\mathcal{I}|}^\top V^\top)^\top  \quad (\because \Tilde{\mathbf{R}} = \mathbf{U}\mathbf{\Sigma} \mathbf{V}^\top)\\
    &= \Tilde{\mathbf{r}}_u (\mathbf{V} \mathbf{\Sigma}^\top \mathbf{P}_{|\mathcal{U}|} \mathbf{\Sigma}_d \mathbf{P}_{|\mathcal{I}|}^\top \mathbf{V}^\top)^\top  \quad (\because \mathbf{U}^\top \mathbf{U} = \mathbf{I}) & \ \\
    &= \Tilde{\mathbf{r}}_u \mathbf{V} (\mathbf{\Sigma}^\top \mathbf{P}_{|\mathcal{U}|} \mathbf{\Sigma}_d \mathbf{P}_{|\mathcal{I}|}^\top)^\top \mathbf{V}^\top
    = \Tilde{\mathbf{r}}_u \mathbf{V} (\mathbf{P}_{|\mathcal{I}|} \mathbf{\Sigma}_d^\top \mathbf{P}_{|\mathcal{U}|}^\top \Sigma) \mathbf{V}^\top \\
    &= \Tilde{\mathbf{r}}_u \mathbf{V} \begin{bmatrix} \mathbf{\Sigma}_d^\top \\ \mathbf{O} \end{bmatrix} \begin{bmatrix} \mathbf{\Sigma}_d^\top & \mathbf{O} \end{bmatrix} \mathbf{V}^\top
    = \Tilde{\mathbf{r}}_u \mathbf{V} \begin{bmatrix} \mathbf{\Sigma}_d^2 & \mathbf{O} \\ \mathbf{O} & \mathbf{O} \end{bmatrix} \mathbf{V}^\top \quad (\because \mathbf{\Sigma}_d^\top = \mathbf{\Sigma}_d) \\
    &= \Tilde{\mathbf{r}}_u \mathbf{Q}^* \text{diag}(1-\lambda_1, \dots, 1-\lambda_d, 0, \dots, 0) \mathbf{Q}^{*\top}  \quad (\because \text{Lemma}\,\ref{lem:b}) \\
    &= \Tilde{\mathbf{r}}_u \mathbf{Q}^* \text{diag}(\underbrace{h(\lambda_1), \dots, h(\lambda_d)}_{\text{Linear Low-pass}}, \underbrace{h(\lambda_{d+1}), \dots, h(\lambda_{|\mathcal{I}|})}_{\text{Cut-off}}) \mathbf{Q}^{*\top}. \quad \quad \quad \square
\end{alignat*}

\begin{lemma}[Eckart-Young-Mirsky Theorem \cite{eckartApproximationOneMatrix1936}]
  \label{lem:a}
  Let $\mathbf{D} \in \mathbb{R}^{m \times n} (m \le n)$ be an arbitrary matrix and $\| \cdot \|_{F}$ be the Frobenius norm. 
  Consider the singular value decomposition of $\mathbf{D} = \mathbf{U} \mathbf{\Sigma} \mathbf{V}^\top$, where $\mathbf{\Sigma} = \mathrm{diag}(\sigma_1, \dots, \sigma_m) \in \mathbb{R}^{m \times m}$ with the singular values $\sigma_1 \ge \dots \ge \sigma_m$. For a given $r \in \{1, \dots, m-1 \} $, partition $\mathbf{U}, \mathbf{\Sigma},$ and $\mathbf{V}$ as follows:
  \[
    \mathbf{U} = \begin{bmatrix} \mathbf{U}_1 & \mathbf{U}_2 \end{bmatrix}, \quad \mathbf{\Sigma} = \begin{bmatrix} \mathbf{\Sigma}_1 & \mathbf{O} \\ \mathbf{O} & \mathbf{\Sigma}_2 \end{bmatrix}, \quad \text{and} \quad  \mathbf{V} = \begin{bmatrix} \mathbf{V}_1 & \mathbf{V}_2 \end{bmatrix},
  \]
  where $\mathbf{U}_1 \in \mathbb{R}^{m \times r}, \mathbf{\Sigma}_1 \in \mathbb{R}^{r \times r}$, and $\mathbf{V}_1 \in \mathbb{R}^{n \times r}$. Then, $\hat{\mathbf{D}} = \mathbf{U}_1 \mathbf{\Sigma}_1 \mathbf{V}_1^\top \in \mathbb{R}^{m \times n}$ is \textit{an analytic solution} of the following optimization problem: 
  \[
    \hat{\mathbf{D}} = \underset{\mathbf{X}}{\mathrm{argmin}} \ \|\mathbf{D - X} \|_{F} \quad \text{such that } \ \mathrm{rank}(\mathbf{X}) \le r.
  \]
  Here, $\|\mathbf{D} - \hat{\mathbf{D}}\|_F = \sqrt{\sigma_{r+1}^2 + \cdots + \sigma_{m}^2}$.
  For $m > n$, the results hold in an analogous manner with appropriate adjustments.
\end{lemma}

\begin{lemma}
  \label{lem:b}
  Let $\Tilde{\mathbf{R}} \in \mathbb{R}^{m \times n}\ (m \le n)$ be the normalized interaction matrix and $\mathbf{L}^{*} \equiv \mathbf{I} - \Tilde{\mathbf{R}}^{\top}\Tilde{\mathbf{R}}$ be the Laplacian matrix of the item-item graph. Consider the singular value decomposition of $\Tilde{\mathbf{R}} = \mathbf{U\Sigma V}^\top$ with well-ordered singular values $\sigma_1 \ge \sigma_2 \ge \cdots \ge \sigma_{m} \ge 0$, and the eigendecomposition of $\mathbf{L^{*} = Q^{*} \Lambda Q^{*}}^\top$ with eigenvalues ordered by $\lambda_1 \le \lambda_2 \le \cdots \le \lambda_n$.
  For the sake of generality, we extend the singular values from $\sigma_1, \dots, \sigma_m$ to $\sigma_1, \dots, \sigma_m, \sigma_{m+1}, \dots, \sigma_n$ such that $\sigma_{i > m} = 0$, referred to as the extended singular values. Note that even if the definition of extended singular vector $\mathbf{v}_{i>m} $ exactly corresponds to $\mathbf{q}_{i>m}^{*}$, it does not result in a loss of generality.

  Let $\mathbf{v}_i$ be an extended singular vector corresponding to $\sigma_i$, and $\mathbf{q}_i^{*}$ be an eigenvector corresponding to $\lambda_i$. Then, the following holds:
  \[
    1 - \sigma_i^2 = \lambda_i, \quad \mathbf{v}_i = \mathbf{q}_i^{*} \quad \forall i = 1, 2, \cdots, n. 
  \]
  Similarly, for $m > n$, the results hold in an analogous manner with appropriate adjustments.
  
\end{lemma}
\noindent \textit{Proof. } Since $\Tilde{\mathbf{R}} = \mathbf{U\Sigma V}^\top$, we have $\Tilde{\mathbf{R}}^\top \Tilde{\mathbf{R}} \mathbf{v}_i = \mathbf{V\Sigma}^\top \mathbf{\Sigma V}^\top \mathbf{v}_i = \sigma_i^2 \mathbf{v}_i$. By subtracting both sides from $\mathbf{v}_i$, $(\mathbf{I} - \Tilde{\mathbf{R}}^\top \Tilde{\mathbf{R}} ) \mathbf{v}_i = (1 - \sigma_i^2)\mathbf{v}_i$. From the definition of the Laplacian matrix of the item-item graph, $\mathbf{L}^{*} \mathbf{v}_i = (1 - \sigma_i^2)\mathbf{v}_i$.
As we already have $\mathbf{L^{*}}\mathbf{q}_i^{*} = \lambda_i \mathbf{q}_i^{*}$ under the given prerequisites, we have $\mathbf{v}_i = \mathbf{q}_i^{*}$, and thus $1 - \sigma_i^2 = \lambda_i$. We have already provided a detailed proof for 
$m \le n$. The complementary case $m > n$ follows vice versa and is therefore omitted. \hfill $\square$

\balance
\bibliographystyle{ACM-Reference-Format}
\bibliography{ref}
\end{document}